# RICERCANDO: Data Mining Toolkit for Mobile Broadband Measurements

Veljko Pejović, Ivan Majhen, Miha Janež and Blaž Zupan

*Abstract*—Increasing reliance on mobile broadband (MBB) networks for communication, vehicle navigation, healthcare, and other critical purposes calls for improved monitoring and troubleshooting of such networks. While recent advances in monitoring with crowdsourced as well as network infrastructure-based methods allow us to tap into a number of performance metrics from all layers of networking, huge swaths of data remain poorly or completely unexplored due to a lack of tools suitable for rapid, interactive, and rigorous MBB data analysis. In this paper we present RICERCANDO, a MBB data mining toolkit developed in a unique collaboration of networking and data mining experts. RICERCANDO consists of a preprocessing module that ensures that time-series data is stored in the most appropriate form for mining, a rapid exploration module that enables iterative analysis of time-series and geomobile data, so that anomalies are detected and singled out, and the advanced mining module that lets the analyst deduce root causes of observed anomalies. We implement and release RICERCANDO as open-source software, and validate its usability on case studies from MONROE pan-European MBB measurement testbed.

## I. Introduction

The advent of mobile wireless communication has had a tremendous impact on numerous aspects of our lives – from the way we navigate in unknown environments, communicate on the move, over the way we pay our bills, to the way we track our health and wellbeing. Underpinning and enabling all of this are mobile broadband (MBB) networks. These networks have witnessed rapid expansion recently – MBB subscriptions have grown more than ten-fold in the last decade and have reached 4.2 billion globally in 2017 [1]. Their performance is improving drastically – a few Mbps download speeds enabled by 3G technology at the break of the millennium appear ancient in comparison with a few Gbps delivered by today's 5G technology. Finally, MBBs are becoming more affordable – worldwide MBB access prices halved between 2013 and 2016 [2]. Together with the expansion of novel paradigms that depend on fast ubiquitous connectivity, such as the Internet of Things (IoT), e-Health, smart cities and factories, the above trends indicate that our reliance on MBB networks is to grow even further.

V. Pejovic, I. Majhen, M. Janez, and Blaz Zupan are with the Faculty of Computer and Information Science, University of Ljubljana, Slovenia. (e-mail: veljko.pejovic@fri.uni-lj.si)

This work is funded by the European Union's Horizon 2020 research and innovation programme under grant agreement No. 644399 (MONROE) through the open call project RICERCANDO. BZ is also supported by Slovenian Research Agency (grant no. P2-0209). The views expressed are solely those of the authors.

Manuscript received DATE; revised DATE.

MBBs inherently have to be very complex to handle global connectivity and a number of different applications for their diverse clients. Networks must seamlessly handle internetworking, user management, accounting, and other aspects of connectivity. At the physical layer MBB networks support different wireless transmission paradigms, such as TDMA, CDMA, OFDM, different modulation and coding schemes, power allocation methods, frequency bands, to name a few technology aspects. Then, at higher levels of organisation, MBB networks have to take care of resource sharing, mobile terminal to base station association, user roaming, and others. Mobile networks have to support different transport (e.g. UDP, TCP) and higher-level protocols (e.g. HTTP, FTP, SIP, etc.), and a wide range of applications – from online social networking, to video conferencing, and augmented reality applications.

The complexity of MBB networks challenges their troubleshooting. Despite the advances in MBB performance measurement methods [3], [4], [5], [6], [7] the problem of the identification of performance anomalies and, even more, the identification of root causes of network anomalies remains unsolved. First, the sheer breadth of networks, both in terms of the number of devices as well as the geography, requires consideration of multiple views of the same phenomenon before any conclusions can be made. Second, the multilayered construction calls for a joint consideration of (meta) information from different levels, from physical layer information on signal strengths, over transport layer retransmissions, to packet delay and jitter. To answer to these needs, MBB network measurement approaches are progressing towards large-scale testbeds capable of providing multifaceted views of network phenomena [8].

Raw MBB measurement data needs to be carefully processed and analysed before any conclusions can be drawn from them. More specifically, mobile data processing pipelines need to 1) detect and remove outliers; 2) consolidate the data coming from different sources, so that they refer to the same event of interest; 3) represent the data in a scalable way that allows examination across different dimensions (e.g. time, space, performance measurements); 4) implement a statistical means for the automatic identification of network anomalies; 5) employ machine learning algorithms to identify factors that might cause the detected network anomalies; and 6) automate data analysis in close synchrony with networking experts and thus support interactive data visualisation approaches.

In this paper we present RICERCANDO, a MBB network data analysis framework developed in tight collaboration of networking and data mining experts and designed to answer to the above-listed requirements. RICERCANDO enables multi-staged and flexible data analysis. Our framework handles the first stage of the analysis through a data representation scheme that merges data of different types and from different sources, and adapts them to time series-based organisation suitable for querying with a different level of granularity. RICERCANDO then enables scalable interactive visual analysis of big network measurement data. Next, to facilitate rapid exploration of the data we implement anomaly detection methods that pinpoint measurements where network performance indicators significantly deviate from the expected values. Finally, we implement a processing pipeline to help with the identification of key factors that might have caused the observed anomalies.

We begin the paper with an overview of the related work (Section II) and then examine the characteristics of MBB measurement data and identify criteria needed for successful network data analysis (Section III). We proceed with the description of the design and implementation of RICERCANDO (Section IV). We then demonstrate the usability of our framework by using it to find anomalies and root causes in data collected by MONROE, a Europe-wide MBB testbed (Section V). In Section VI we present the most important lessons we learnt while developing and using RICERCANDO. We have released RICERCANDO as open-source software and in Section VII we present concluding remarks and at the same time invite the community to join our efforts towards supporting rapid MBB network measurement data analysis.

## II. RELATED WORK

A systematic means of monitoring is crucial for assessing the quality of service and troubleshooting in mobile broadband networks. Recently, a wide range of approaches for MBB measurements have been developed [9]. Approaches rely on either passive [4], [7] or active measurements [10], or on a hybrid measurement methodology that combines both [11], [12], [13], [3]. Passive measurements merely observe the existing network traffic, while active measurements inject own packets in order to evaluate performance metrics. The downside of active measurements is that the measurement process may impact the actual network under test. In terms of the measurement point locations, certain approaches, especially those initiated by national regulators, use dedicated monitoring equipment and a small number of controlled nodes, while others rely on crowdsourced measurements conducted by a large number of often uncoordinated users [14]. The former have the benefit of being unrestricted by the provider, of viewing the network "as users", and of covering wide geographical areas. OpenSignal, for instance, has more than 100 million users across the globe [15]. However, crowdsourced measurements suffer from unreliability due to the lack of control over the measurement equipment. A mobile app-based measurement software may be run on different phone models, with different implementations of the operating system, with devices running different applications in parallel to the measurement app, with different hardware issues (e.g. bent antennas), and with devices placed in various locations during measurements (e.g. bag/pocket/hand), all of which may impact measurement results [16], [17]. Recent commercial and research initiatives hence use crowdsourced-like approach with specialised equipment dedicated to network measurements [8].

Irrespective of the measurement approach, MBB measurement data is large-scale, temporal, heterogeneous, and shaped by a number of factors related to measurement methodology and equipment. Storing, processing and reasoning upon such data is challenging, and a number of solutions to assist with the above tasks have been developed. Svoboda et al. demonstrated the importance of using a well-defined methodology for delay measurement to obtain meaningful interpretation of the results [11]. CoMo enables fast prototyping of network measurement mining applications [18]. However, mostly concerned with data storage and flow, CoMo does not provide sufficient support for advanced analytics. In addition, CoMo focuses on TCP and UDP and does not consider measurements at the lower MAC and PHY layers, crucial for troubleshooting MBB networks. Future efforts were aimed at either increasing scalability, usability, or the number of supported options for data analysis. ENTRADA, for instance, converts `pcap` log file to Apache Parquet and enables stream mining [19]. Similarly, DBStream was built to support rolling big data analysis [20]. The tool's utility has been demonstrated on a few use cases, including on the analysis of signalling and data transfer behaviour of different mobile device types and different operating systems [21]. However, designed by networking experts, these systems usually provide solutions to network measurement data handling, yet stop at the point where advanced data mining is needed.

Consequently, networking researchers often resort to ad-hoc approaches to data mining. Baltrunas et al. show that even simple correlation can help with network reliability estimates [22]. In order to profile network coverage in Norway, Lutu et al. perform hierarchical clustering of measurement data collected via train-mounted probes [23]. Narayanan et al. go a step beyond off-the-shelf data mining approaches and propose a feature distribution similarity graph to analyse spatio-temporal mobile measurement data [24]. The authors show the utility of the approach in a case study of profiling mobile users' behaviour from call detail records.

More advanced approaches try to automate the mining process, especially when it comes to anomaly detection, a key issue in network data analysis. ADAM system detects anomalies by calculating the Kullback-Leibler divergence between the incoming and previously collected data [3]. Once an anomaly is detected, the system performs factor analysis to identify features exhibiting a similar abrupt change. RCA tool initially detects change points by measuring the entropy of considered features. It then considers the full statistical distribution of the traffic features to characterise anomalies [25]. Ricciato et al. suggested two approaches to bottleneck detection, the first one based on statistical analysis of the aggregate rate, and the second method based on TCP performance indicators [26]. Coluccia et al. proposed an anomaly detection methodology that identifies statistically significant deviations from the past behaviour using Maximum Entropy modelling [27]. In another study, the authors investigated distributions of multiple features to detect traffic anomalies, indicating that the alarm correlation across features may augment the accuracy of the detector [28]. Other statistical methods for anomaly detection were further proposed [29]. Unique to all approaches is that they require a networking expert in the loop. This is explicitly evident in Siekkinen et al. TCP RCA approach [30], but also through subtle issues related to data collection and interpretation process. For instance, Michelinakis et al. show how peculiarities of packet scheduling at an LTE base station impact capacity estimates inferred through measurements [31].

## III. MOBILE BROADBAND MEASUREMENT DATA CHARACTERISTICS AND DATA ANALYSIS REQUIREMENTS

In this work we develop RICERCANDO, a generally applicable framework for mobile broadband measurement data mining. The framework was built in collaboration with MONROE [8] – an open access hardware-based platform for independent, multihomed, large-scale experimentation in MBB networks. RICERCANDO readily supports data formats used by the MONROE project (other data sources might require minor adjustments).

The MONROE project aims to create a pan-national reliable open-access measurement platform for MBB networks[1]. The core of the system is a MONROE node, a custom-built device fitted with a Debian-based single board computer and up to three LTE modems connected to different providers. A centralised experiment scheduling system allows MONROE users to post custom-made experiments to distributed nodes and remotely collect the measurement results. In addition, each node independently executes certain background experiments, such as periodic RTT measurements to MONROE servers. Finally, all the experiment data and meta-data are collected in a MONROE database implemented in Cassandra[2]. In 2018, the project operated 150 measurement nodes in four European countries, with more than a half of the nodes being mounted on buses, trains, and delivery trucks.

MONROE data, similarly to other MBB measurement data are characterised by:

- Spatio-temporality: measurement nodes are mobile;
- Multi-modality: multiple aspects of network performance (RTT, throughput, etc.) and meta-data (location, CPU load, etc.) are sampled in parallel;
- Heterogeneity of data granularity and the lack of synchrony among different measured features;
- Impact of the measurement methodology, hardware, and software on the measurement results;
- Lack of ground truth data.

On the implementation level, a MBB data analysis tool has to cope with the above characteristics of the data. On the higher level, the tool has to enable comprehensive analysis, requirements of which have been discussed among the research community before. For instance, in 2006 Ricciato indicated that network traffic analysis should include statistical analysis that goes beyond simple ad-hoc solutions, visualisation and multidimensional exploration by networking experts, advanced machine learning modelling algorithms, and should allow the data to be pipelined to other tools [32].

We design RICERCANDO after a careful consideration of the existing work, open issues related to network measurement analysis, and guidelines from the community [32]. In RICERCANDO, we explicitly support interactive analysis and put the user in the loop. Moreover, our data storage paradigm is adapted to support rapid visualisation and experimentation, so that the expert knowledge can be harnessed in the best possible way. Similarly, identifying a need for automated statistical analysis, we create a machine learning pipeline that automatically detects and suggests explanations for network anomalies. At the same time, the system's visual component maintains a close dialog with an expert enabling iterative investigation until the root cause of the issue is identified. Finally, recognising the uniqueness of each measurement setup and varying goals of those who analyse networks, we do not restrict RICERCANDO to particular mining techniques. Rather, we integrate it with the popular data mining suite Orange[3], allowing a wide range of current and future data mining approaches.

## IV. RICERCANDO FRAMEWORK

RICERCANDO is structured around modules that together create a data mining pipeline (Fig. 1). The framework assumes that the data is stored in a key-value database, such as Cassandra used by the MONROE project. *Data Preprocessing* module transforms and stores the data so that it can be quickly retrieved along the temporal dimension. *Data Acquisition Interface* enables different views over the data. *Rapid Exploration* module consists of three submodules that allow interactive visualisation of time-series data, geomobile data visualisation, and rapid anomaly detection in the data. Finally, *Advanced Mining* module interfaces with Orange data mining suite and enables sophisticated machine learning and data visualisation methods.

RICERCANDO software implemention consists of a core `ricercando` Python library[4], data preprocessing scripts written in Bash and Python, Jupyter Notebooks for visual analysis of the data, and an add-on for Orange data mining suite. All the code, together with the installation instructions is available on GitHub[5].

### A. Data Preprocessing and Interfacing

MBB measurement data are often collected in relational or key-value databases, as they enable easy and efficient storage [33], [20], [3]. However, stored in such a manner, data are not suitable for rapid interactive exploration. This is especially true for data having a temporal dimension, which is common in MBB measurements – nodes move in space/time, RTTs are gathered with periodic pings, anomalies and glitches impact subsequent node behaviour, to name a few time-related aspects of the measurements. Key-value and traditional relational databases severely limit the performance and the flexibility of writing queries over time-series data. Second, the volume of data and metadata gathered by MBB measurements can be large. For instance, RTT measurements from MONROE platform produce approximately 20 million entries per day. Data storage needs to support data sampling to allow zooming in and out on a selected chunk of data, or to support concurrent analysis of data coming from multiple nodes. Finally, MBB data comes from various sources – multiple nodes and multiple processes within a measurement node – and are often not aligned along the common time axis. Consequently, merging the data in order to enable multidimensional analysis is challenging.

---

[1] http://www.monroe-project.eu
[2] http://cassandra.apache.org
[3] http://orange.biolab.si
[4] `ricercando` is also available via `pip` installer
[5] http://github.com/ivek1312/ricercando/

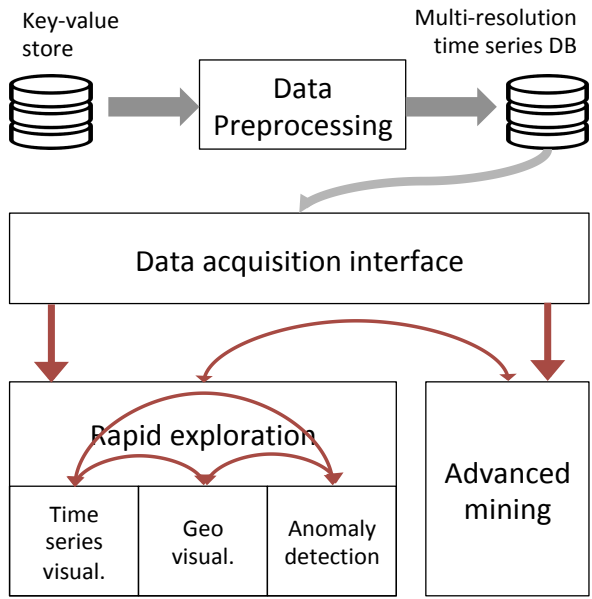

Fig. 1. An overview of RICERCANDO framework. Boxes represent framework's modules, while arrows represent data movement. Darker (red) arrows indicate that data is given in Python pandas format, suitable for interchange among different data processing modules and tools.

In RICERCANDO we devise data transformation and data storage schemas to transform MBB data into minable representations. We use temporal data abstraction and feature engineering guided by domain-specific knowledge, and we construct scripts that implement various data transformation tasks. To solve the temporal data mining problem we transform the data to a time-series database[6]. We store time-series data with the minimal temporal granularity (10 ms for the MONROE data use case). Furthermore, we also sample and store the data at different granularities (1 s, 1 min, and 30 min for the MONROE data use case). This is crucial for enabling interactive visualisation – if a user requests to visualise a whole day of data, we fetch data of a coarser temporal granularity; for examining particular anomalies, we zoom in and provide fine grain data. When sampling to low resolution the aggregation of values within the period depends on the type of data. Thus, with a few exceptions, for categorical variables we use *mode* function that returns the most frequently observed value in the considered time frame, while for numerical we use either *min*, *max*, or *mean*, depending on the particular variable.

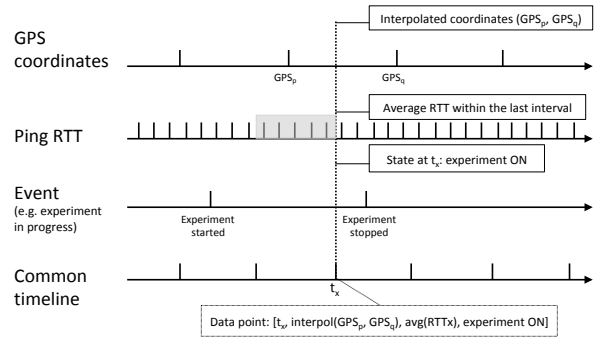

Fig. 2. Data merging along the common time axis in RICERCANDO.

Data mining and modelling is performed on datasets consisting of *instances*, where each instance represents a data point in a multidimensional feature space. For example, a measurement of the GPS location, RTT, and the state of the measurement node at a point in time. As measurement data come from various non-synchronised sources, we often have to merge individual data streams along the same time axis. A sketch of the merging process we implement in the Data Acquisition Interface module is shown in Fig. 2. For each of the time series (e.g. ping RTT, GPS coordinates, etc.) we find an intersection with a selected moment on the common time axis. Depending on the nature of the feature described by the time series we apply a different strategy for getting the value at the requested moment in time. For instance, for GPS coordinates we perform interpolation between the last measurement before and the first measurement after the given moment in time. For RTT we take an average of the measurements recorded in a time window preceding the current moment. For the feature indicating events on a node we keep track of the node's state – for example, whether an experiment is currently running at the node. Finally, RICERCANDO allows further tuning of the merging process, for instance, by specifying the minimum freshness value of the data before it is included in a data instance – e.g. if no download speed measurements were taken in the last 60 s, the instance will contain a *null* value for *download speed*. While we steer away from a fully automated merging and require input from a networking expert, this guarantees that the further analysis is done on truly meaningful data[7].

### B. Interactive Visualisation of Big MBB Measurements Data

Iterative examination of visualised data is crucial for network data mining [32]. These data, however, are multidimensional, temporal, and geo-mobile, and very large, thus inappropriate for analysis using conventional data visualisation tools that come with data mining packages, such as WEKA or Orange [37], and they might even overburden specialised tools, such as Tableau [38]. In RICERCANDO we develop two modules for rapid interactive visualisation of MBB measurement data – one for time-series visualisation, the other for geographical data visualisation, both implemented in the form of Jupyter Notebooks. We opted for this environment, as opposed to custom stand-alone programs, as it allows quick prototyping and tweaking according to specific user needs and given datasets.

---

[6]We use InfluxDB (www.influxdata.com) in our implementation, other time-series databases are also appropriate.

[7]The inclusion of domain experts early on in the data preprocessing stage is often emphasised as a crucial step in modern data mining [34], [35], [36].

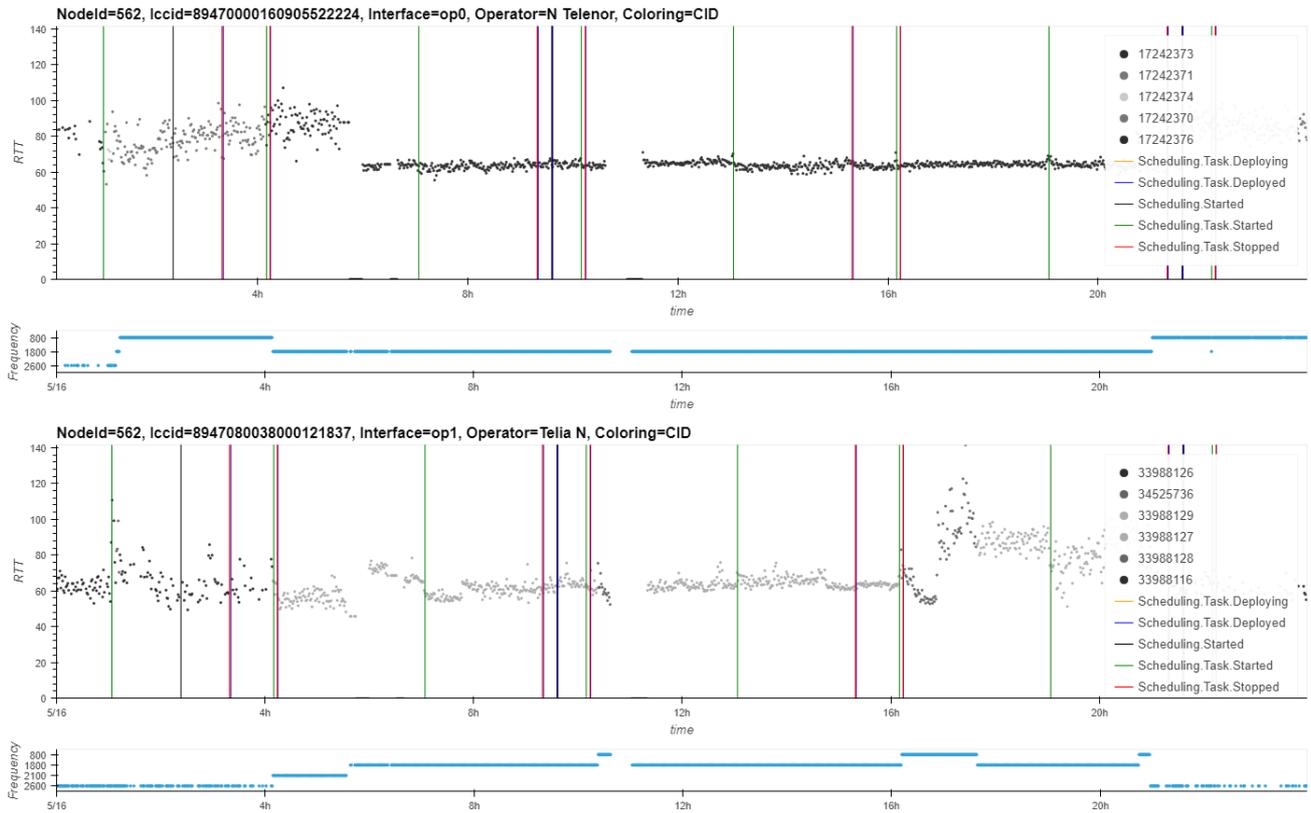

Fig. 3. Time-series visualisation in RICERCANDO. Y-axis represents RTT measured on each of the two interfaces of the same node, while colouring corresponds to the cell id (CID). Vertical lines represent MONROE experiment start/stop/loading moments. Plots below each of the RTT series show the frequency used by the interface.

**Time-Series Visualisation module**, for a selected network probe (node) and a time period, plots a target key performance indicator (KPI) on a separate timeline for each of the node's interfaces. An additional dimension can be represented through the colouring of each of the points (Figure 3). Finally, hovering over a point shows values of all the other dimensions associated with the same data point. A key property of the Time-Series Visualisation module is its adaptability to the amount of to-be-shown data. It relies on `getdf` function from `ricercando` Python module, which, for the given zoom level retrieves data from the database with an appropriate resolution, in order to preserve the interactivity of the notebook. For example, viewing a whole week worth of measurements might use data aggregated on 30 min intervals, whereas zooming into a particular RTT anomaly might fetch and show data with 10 ms granularity.

**Geographical Data Visualisation module** (Figure 4) supports visualisation of a selected KPI of geo-referenced data from a measurement node on a separate map for each of the node's interfaces, for the given time period. Such visualisation is a key tool for the identification of issues affecting particular geographic regions. Similarly to the Time Series Visualisation module, hovering over a point shows values of all the other dimensions associated with the same data point. Geographical Data Visualisation module, too, relies on `getdf` function for adaptive data retrieval, so that the retrieved data resolution is adjusted to the current map zoom level.

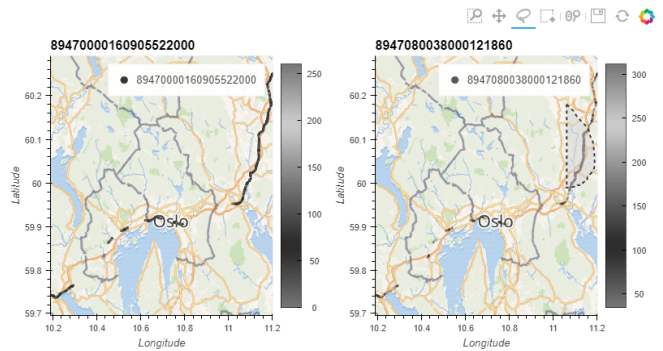

Fig. 4. Geographical visualisation of RTT measured on two interfaces of the mobile node travelling through Oslo. The shades of the trace correspond to different values of RTT. On the right image a selected region contains RTT data stored for further analysis.

RICERCANDO modules, such as Time Series Visualisation, Geographical Data Visualisation, Anomaly Detection, and Advanced Mining module are designed to fit into each other just like LEGO® bricks and allow flexible data analysis workflows. To support interoperability among modules we rely on Python pandas DataFrame (dark/red lines in Figure 1). Indeed, each of the Jupyter notebook-based modules allows data selection (e.g. selecting a range of data points on a map) and storage (as a DataFrame on local storage), and retrieval from another module, e.g. in order to perform advanced mining in Orange.

*C. Anomaly Detection Tool*

Anomalies occur frequently in computer network measurement data and can be caused by anything from misconfigurations to cyber attacks [39], [40]. Anomaly detection plays a central role in RICERCANDO. We implement a Jupyter Notebook that enables automatic detection and visual inspection of anomalies in the data (Figure 8). Numerous detection methods relying on a range of techniques, from mining association rules [41], to modelling with Markov processes[42], have been proposed for anomaly inference (see [43] for a survey of anomaly detection approaches). In RICERCANDO we implement three anomaly detection methods:

- **Rolling mean** – a method based on a rolling window that compares data in the current window with the long-term mean of the measurements. Data points that are a number of standard deviations away from the rolling mean are regarded as outliers and a large enough cluster of outliers is deemed an anomalous region. With this method abrupt falls or rises (spikes) are treated as anomalies, while, for example, a gradually rising RTT due to increased network congestion would not be considered an anomaly.
- **Baseline comparison** – a detector that compares the actual value of a data point with the value predicted based on a pre-constructed model. Such a method can, for example, learn the expected RTT for a node using 4G technology experiencing a certain RSRQ in a certain area, and correctly attribute changes in RTT to either contextual changes – like fallback from 4G to 3G – or to an unexplained anomaly. Due to a large parameter space the observed data point might come with a previously unseen context. To cope with such a case, RICERCANDO builds the model using the quantile regression forest technique [44] that predicts the dependent variable value even if the context has not been observed before. Furthermore, we build a model by using top N (by default 10) percent of the best performing measurements from a given context. This ensures that well performing points are not misclassified as anomalies.
- **Distribution comparison**[8] – a detector that empirically infers distributions of the same variable in different segments of the data using kernel density estimation technique, and then compares the distributions using Kullback-Leibler divergence. Significant difference between the previous and currently observed data distributions may indicate an anomaly.

> **What is an anomaly?**
> Without any knowledge of the underlying system that generates the data, an anomaly detection system aims to find "sufficiently different" measurements in a stream of data. Alternatively, the data are labelled as "anomalous" if they do not follow the patterns that a domain expert expects, based on her mental model of how the MBB network "should" behave. While the first definition leaves us struggling to find parameter values that would define "sufficiently different" behaviour in automated anomaly detection systems (Romirer and Ricciato have pondered on this question in the context of delay measurements in 3G networks [45]), the second definition is limited by the expert's (mis)understanding of the network phenomena. Thus, in RICERCANDO we aim to judiciously guide an expert in reasoning about the observed deviations. The rolling mean and distribution comparison methods we implement allow automated labelling of "sufficiently different" measurements, while the methods' parameters allow the experts, guided by an immediate visual feedback, to adapt the labelling to the situation at hand. Baseline comparison method, on the other hand, "encodes" the knowledge learnt on previously seen data, and labels as anomalous only those values that do not conform to the constructed model, therefore, moving the automation closer to the "expert" side of the spectrum.

The developed notebook allows the user to select a measurement node, a target KPI, and a time span in which the data is analysed. Additionally, the user can set a number of parameters that control the operation of the tool, i.e. the sensitivity of anomaly detection. In the first step the developed tool automatically detects the anomalies in measured data based on one of the above detection methods selected. Besides these methods, the tool supports simple integration of new anomaly detectors. After one or more anomalies are detected, the tool enables informative visualisation of regular and anomalous data. Based on visual results a domain expert may adjust initial parameters to control the shape of the highlighted anomalies. This is demonstrated in Figure 5, where tuning of parameters produced two different anomaly regions within the same data. Descriptive visualisation also allows the experts to quickly find important aspects in the data. The data can then be saved so that anomalous regions are automatically labelled for further processing.

An important feature of the anomaly detection tool is concurrent anomaly detection. MBB data often contains measurements from a large number of nodes connected to a few different network providers, and detecting anomalies that simultaneously appear at all interfaces connected to the same provider is crucial for identifying whether the anomaly is isolated or affecting the whole network. In RICERCANDO we implement an optional concurrent analysis that takes into account all probes connected to a particular network. The output of the tool is a time diagram showing a cumulative count of anomalies over time for the selected network – moments when such a count is high indicate network-wide issues (see Section V-D).

---
[8]This method is not suitable for "running" data analysis, therefore, we implement it in the Anomaly Detection module, but do not expose it through our GUI.

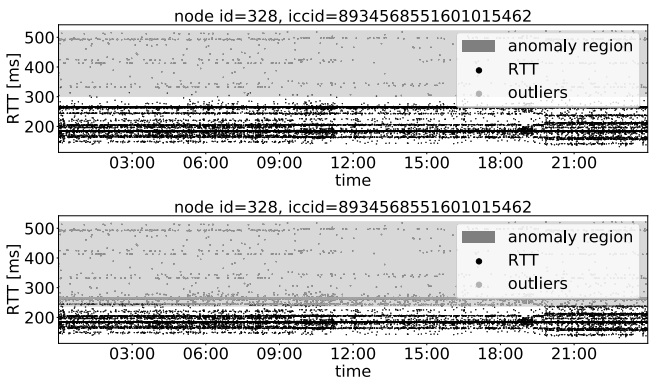

Fig. 5. Anomaly detector determined two different anomaly regions (higher RTT values shaded grey) within the same data by using distinct values of detection parameters in each case.

### D. Advanced Mining

Identifying root causes of the observed MBB behaviour is the final goal of data analysis. The existing tools for MBB data analysis were mostly developed by computer networking experts and support data preprocessing, visualisation, and simple statistical analysis [18], [30], [22], [3]. RICERCANDO is developed in close collaboration with highly experienced data mining experts – one of the RICERCANDO authors is leading a 19-member data mining research lab with more than 20 years of practical data mining experience in a range of domains. This synergy enables us to support advanced data mining for root cause analysis in RICERCANDO.

A key enabler of advanced mining in RICERCANDO is Orange – a popular GUI-based data mining toolbox where data processing workflows are constructed through visual programming by combing *widgets*. A widget is a computational unit with interactive visual interface that performs a particular function related to data preprocessing, visualisation, and modelling. Orange supports a range of machine learning methods, from unsupervised (clustering), to supervised (classifiers, regressions), from basic (e.g. naive Bayesian) to more complex state-of-the-art ones (e.g. neural networks). Figure 6 depicts MONROE measurement data analysis using an Orange workflow of widgets.

However, Orange is limited in the amount of data it can handle. Thus, we use it as the last step of RICERCANDO analysis. We develop an Orange widget to import the data from RICERCANDO rapid exploration notebooks. Users can, thus, perform preliminary visualisation and analysis of a larger dataset in a Jupyter Notebook before selecting a particularly interesting dataset and analysing it further in Orange. In addition, we develop a widget for direct access to MONROE data stored in a time-series database.

One of the main questions a network analyst is interested in is *which factors may cause a particular anomaly?* [46].

To answer this, we develop an Orange widget that identifies *Significant Groups* of features that differentiate between regular and anomalous data. Note that a dataset containing labelled regular and anomalous data is automatically created by our Anomaly Detection module and imported to Orange via the iPython Connector widget. The main test implemented within the Significant Groups widget is the hypergeometric test. The test traverses all subsets of features and calculates the enrichment each subset brings to the anomalous data region. Sorting the subsets according to the enrichment, while also taking into account their significance levels, gives us a list of most probable causes for the detected anomaly. In addition, the widget supports other comparison tests that may help with root cause analysis, such as the permutation test and the $t$-test.

## V. CASE STUDIES

The MONROE project provides large amount of data of various MBB network parameters. Irregular patterns in the data can quickly be spotted using the visualisations. However, to precisely define visually observed anomalies and to discover hidden anomalies that are not easy to illustrate, we developed an automatic anomaly detector. Beside identification of anomalies the computer tool also facilitates the determination of their root causes. Among multiple occurrences of anomalies that we found, selected case studies focused on RTT data are thoroughly described in this section.

### A. Connection Mode Change

The first "anomaly" we identified by using our automatic detection tool's rolling mean method is depicted in Figure 7 (top). The figure shows how after 11:30 the RTT mean changes drastically from below 100 ms to approximately 250 ms. The anomaly detector automatically recognised the shift and marked it as an anomaly (grey region). Running the hypergeometric test and calculating the enrichment each feature subset brings to the anomalous data region, we found out that a change in the device's connectivity mode is the culprit. A switch from LTE to 3G perfectly coincides with the anomaly, as shown in Figure 7 depicting the RTT and the interface's mode on the common time axis. Note, however, that by automating the significant feature search we remove the need for comprehensive visual analysis.

This initial example shows the limitations of the automated approach relying on domain-agnostic data deviation detection (see discussion in Section IV-C). In Section V-C we present a model-based approach, which, armed with the knowledge based on the previously seen data, correctly considers the above example to be non-anomalous, as it can be easily explained through the network interface mode change.

### B. Measurement System Interference

In many instances we encountered sudden short-lasting drops in the measured RTT. Figure 8 shows RTT measurements within two hours from 20:00 to 22:00 on one of the interfaces. The majority of measurements have values near 100 ms, but between 21:05 and 21:20 there is a concentrated group of measurements with values around 80 ms. The shaded area marks an anomalous group which was identified by our rolling mean detector. Many dispersed outliers can be seen in Figure 8, yet only a cluster with a sufficient number of outliers composes an anomaly. In such situations the detection using the computer tool is more accurate than just a visual observation of data.

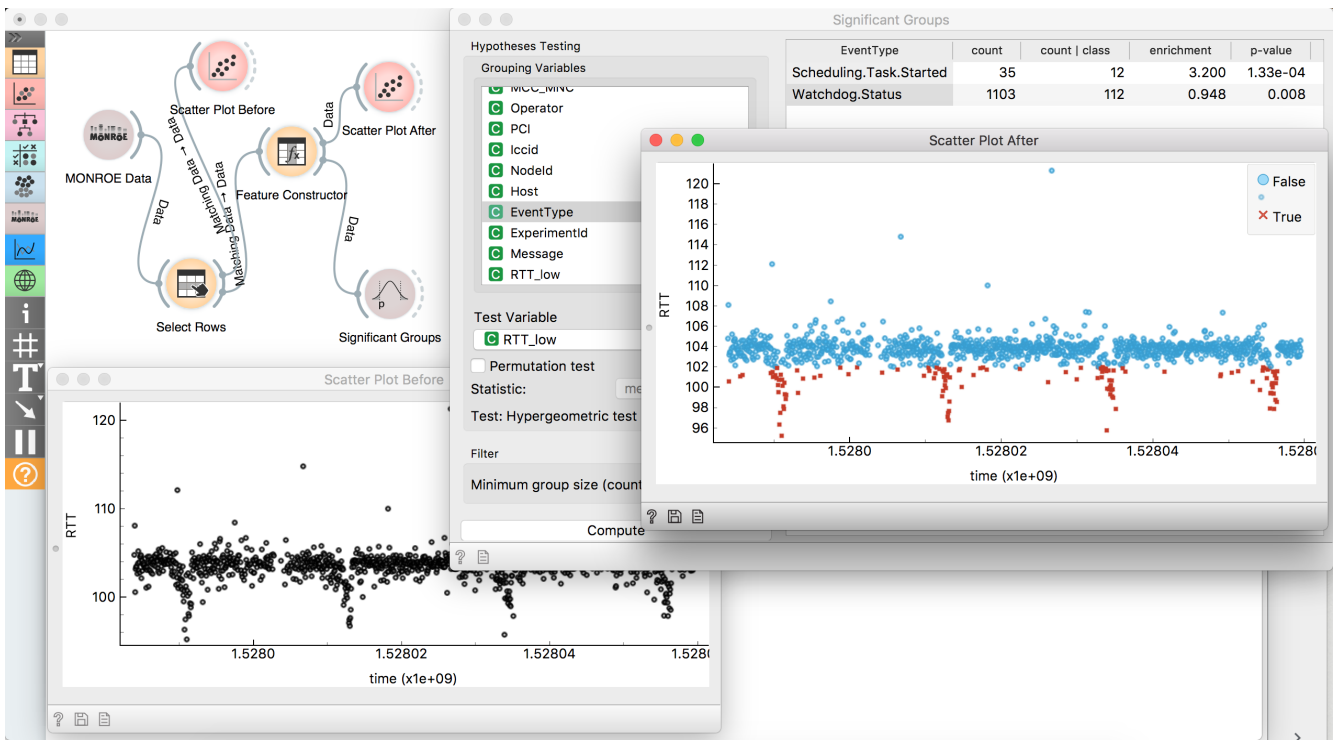

Fig. 6. MONROE data analysis in Orange. A workflow composed of Orange widgets is shown in the upper left corner. Each widget performs a specific function. A window corresponding to *Scatter Plot Before* widget (lower left) shows anomalous RTT behavior. *Scatter Plot After* window (middle right) shows distinct RTT "dips". *Feature Constructor* widget is used for splitting the data into groups with low and normal RTT. Finally, *Significant Groups* widget performs a hypergeometric test and identifies `Scheduling.Task.Started` event as a feature value that discerns between the two groups, indicating that a background experiment impacts the observed RTT.

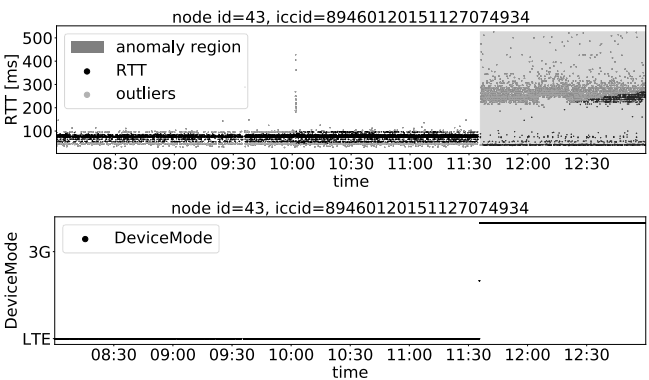

Fig. 7. The image on top shows the increase of RTT measurements after 11:30 and the bottom image shows how the shift correlates with the change of parameter DeviceMode. The vast majority of RTT values before 11:30 are around 100 ms, so the relatively rare outliers at that time do not form an anomaly.

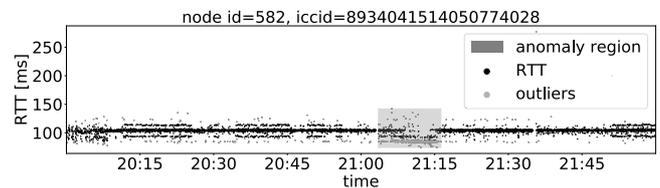

Fig. 8. The anomaly detected via rolling mean method is marked with a shaded area. Occasional outliers are coloured grey, however, they do not necessarily compose an anomaly.

The significance analysis for this case shows that the root cause of this anomaly is the event `Scheduling.Task.Started` (Figure 9), indicating that a start of an experiment on a node causes the anomaly. It seems that running an experiment on a node triggers a drop in measured RTT values.

We hypothesise that the cause of such behaviour is the discontinuous reception (DRX) mode. DRX allows interfaces to save energy by going to a low power mode when no data is being transmitted [47]. However, DRX may lead to the RTT increase if the ping packets, before the transmission, have to wait for the interface to go back to a high power state. MONROE platform pings are sent out with 1 second inter-packet time, while operators often set the DRX kick-in threshold to around 100ms. Consequently, we expect that most of the MONROE ping packets, unless an interface is already active because of an experiment, indeed have to wait for the interface to go to the high power state before the RTT measurements can be performed. To confirm the existence of DRX we conducted our own ping experiments on the MONROE platform with variable inter-packet times. As expected, once ping packets were sent out with a higher frequency, the measured RTT dropped.

| Iccid,Variable | count | count \| class | enrichment | p-value |
|---|---|---|---|---|
| 8934041514050774028,EventType=Scheduling.Task.Started | 1791 | 366 | 1.15067 | 7.0252e-21 |
| 8934041514050774028,Host=130.243.27.221 | 6289 | 372 | 1 | 0 |
| 8934041514050774028,Operator=YOIGO | 6289 | 372 | 1 | 0 |
| 8934041514050774028,IP_Address=10.41.236.41 | 6289 | 372 | 1 | 0 |
| 8934041514050774028,Band=3 | 6287 | 372 | 1 | 0 |
| 8934041514050774028,CID=72209510 | 6289 | 372 | 1 | 0 |
| 8934041514050774028,DeviceMode=LTE | 6289 | 372 | 1 | 0 |
| 8934041514050774028,DeviceState=connected | 6287 | 372 | 1 | 0 |
| 8934041514050774028,Frequency=1800 | 6289 | 372 | 1 | 0 |
| 8934041514050774028,DeviceSubmode=unknown | 6287 | 372 | 1 | 0 |
| 8934041514050774028,LAC=65535 | 6287 | 372 | 1 | 0 |
| 8934041514050774028,PCI=65535 | 6287 | 372 | 1 | 0 |

Fig. 9. Significance analysis determines the event `Scheduling.Task.Started` as the root cause of the anomaly shown in Figure 8.

### C. Baseline Model Anomaly Detection

The rolling mean anomaly detection method is limited in its ability to adapt to well understood changes in the observed variable. For instance, in the case examined in Section V-A the jump in ping RTT measurements is not unusual, having in mind the device connection mode change. On the other hand, the baseline method for anomaly detection uses a pre-constructed quantile regression tree model to infer the expected value of the observed parameter in the light of the given context, i.e. values of selected remaining parameters. Consequently, the method does not mark as anomalous those measurements that can be explained with the pre-constructed model. This greatly reduces the number of false positives, as an "anomaly" can, in fact, be explained by the model.

In Figure 10 we show model-predicted values of ping RTT (black line) and the observed values (grey dots). The baseline model takes in to account RSSI (Received Signal Strength Indicator), RSRQ (Reference Signal Received Quality) and RSRP (Reference Signal Received Power) as independent variables. The prediction was created by quantile regression forests algorithm, taking into account the top 10th percentile of predicted RTT values. Here the higher percentile indicates better/smaller RTT value. The outliers are the points distant from the baseline, meaning that their actual value highly disagrees with predicted value. In top image are two shaded anomaly regions formed by outliers high above the baseline.

The first anomaly can be explained by further refining the model with the information on the expected RTT at different cell IDs (CID) that a device connects to. This is clarified in bottom image in Figure 10, where the baseline was constructed by quantile regression forests algorithm that predicts RTT values with respect to CID. The three steps in the baseline function in the bottom image correspond to three different cells that a device connected to. Therefore, the first anomaly constructed in top image is not considered an anomaly, if the CID parameter is taken into account. Note that the figure still does not explain why RTT measurements differ across different CIDs – this requires further investigation that goes beyond the capabilities of the collected dataset. The second anomaly on the right side of both images in Figure 10, however, is not due to different CID, so its root cause is the variation in parameters other than CID, RSSI, RSRQ, or RSRP. In this way the baseline anomaly detector not only uncovers anomalies that are impossible to detect visually, but can also explain anomalies by choosing the appropriate independent variables for the quantile regression forests algorithm.

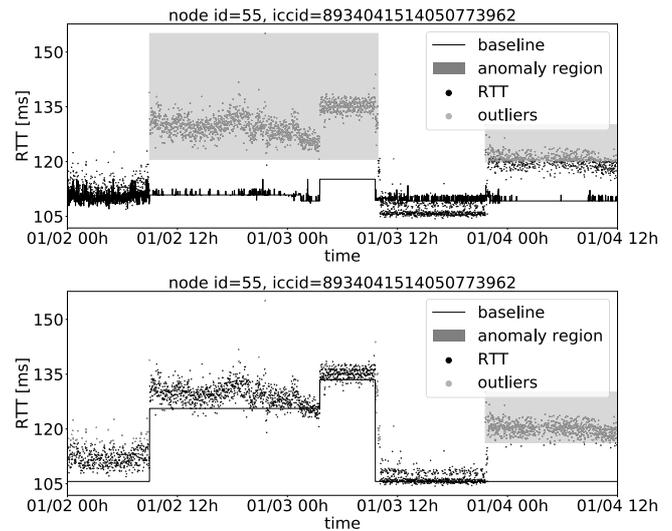

Fig. 10. In the image on top the baseline is the predicted value of RTT with respect to parameters RSSI, RSRQ, and RSRP. The detector marked two anomalies. The first anomaly is resolved by using CID to construct the baseline as shown on bottom image.

### D. Network and System Wide Anomalies

We are further interested to determine whether a certain anomaly appears only at a particular network interface or, perhaps, at a number of interfaces connected to the same Internet service provider (ISP), or even beyond – in a number of devices across the measurement system. Such case could indicate systemic causes of the anomalies. In order to study such examples we enhanced our anomaly detection tool to support concurrent anomaly detection over a number of interfaces – essentially, it counts all anomalies happening at the same time at nodes connected to the same ISP. Figure 11 shows the number of anomalies that occurred simultaneously at all nodes connected to ISP YOIGO on June 3rd 2018. A pattern of periodic spikes can be observed. This anomaly is due to an RTT drop caused by a running experiment, similarly to the case examined in Section V-B. The large number of concurrent anomalies at spikes correspond to experiments scheduled to run on different nodes of the same operator at the same time.

We further examined potential network-wide anomalies. Through exploratory analysis at a few interfaces we noticed an anomaly caused by missing data. We then ran the concurrent anomaly detection tool for all the interfaces connected to a few different ISPs. In Figure 12 we show the cumulative anomaly count for two different ISPs – Vodafone IT and YOIGO. We see that both operators exhibit simultaneous peaks that are more than two standard deviations above the mean anomaly count. The same peak is observable with other ISPs (not shown in the figure). This indicates a system wide anomaly, likely caused by a glitch in the measurement system.

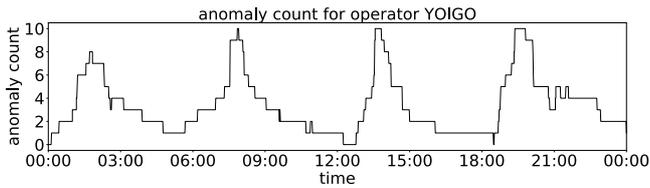

Fig. 11. Number of simultaneously occurring anomalies at all nodes connected to the same ISP.

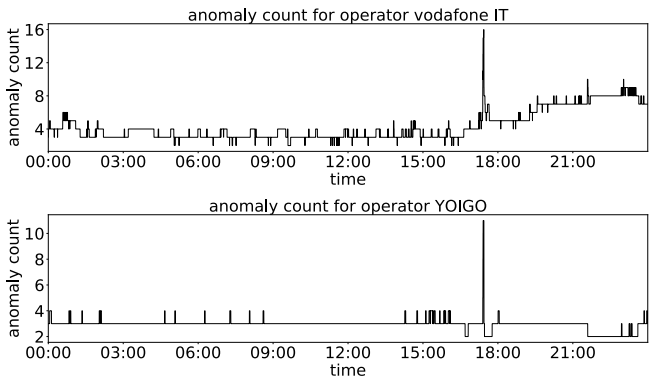

Fig. 12. A system-wide anomaly due to the missing RTT data at approximately 17:30 on January 1st 2018.

## VI. LESSONS LEARNT

Continuous experimentation and revising has marked the process of RICERCANDO design and development. Different prototypes have been developed, applied on the data, and evaluated, while at the same time the underlying measurement platform (MONROE) kept evolving, essentially making our goal a moving target. In this section we present some of the main lessons learnt through the development process.

*Need for appropriate data preprocessing and representation.* At the time RICERCANDO started in June 2016 the MONROE platform was producing only a modest amount of (meta) data from a limited number of nodes. However, over the course of the project the amount of collected data grew both because additional nodes were deployed, as well as because of the additional information that was collected on each node (e.g. different background experiments). MONROE data are by default stored in a Cassandra no-SQL database. This, however, severely limits large-scale data mining of the platform data. While Cassandra enables easy storage of key-value pairs, it is inappropriate for mining temporal data. Most of the collected data indeed have a temporal dimension, thus time-based querying remains crucial. Another issue with Cassandra is that it does not support data sampling. In MONROE, data are often collected with very fine granularity (e.g. a ping every second), which makes (visual) inspection over a larger time period impractical – there are simply too many points to be shown on a graph. In the early stages of RICERCANDO we tried to adapt to the given database. However, in the next step, in order to enable efficient temporal large data analysis we devised a solution that relies on InfluxDB, a database specifically targeting time series data querying.

Joining tables over the common timestamp field is another challenge we have faced. Since timestamps are asynchronous, some tolerance on timestamp joining had to be accounted for. One solution was sampling data at rounded timestamps directly on the database, which we also used for visualisation. Another solution was provided by pandas library – `mergeasof`, function similar to a left-join except that we match on nearest backward timestamp with defined time tolerance. This helped us obtain more meaningful data point instances with fewer missing values. Data preprocessing and representation is usually the most difficult step, especially when dealing with large amounts of data. Our contribution, released in a form of processing scripts makes this step a lot easier for MONROE platform users.

*Available data imposes explanation capacity limits.* The interpretation of some encountered anomalies eluded us. One of these is depicted in Figure 13. A drop in mean RTT value occurs around 7:00, similar to the case of ping experiment running on the node (Figure 8). However, there were no scheduled experiments in the case in Figure 13, so they are ruled out as root-cause of the anomaly. Also, the 2-hour extent of this anomaly is longer than the 10-minute duration of an experiment. Furthermore, the anomaly appeared only at one interface of the same node. The available data is simply insufficient for explaining this anomaly and more information, perhaps those coming from the specific logs of operation of this particular device, is needed.

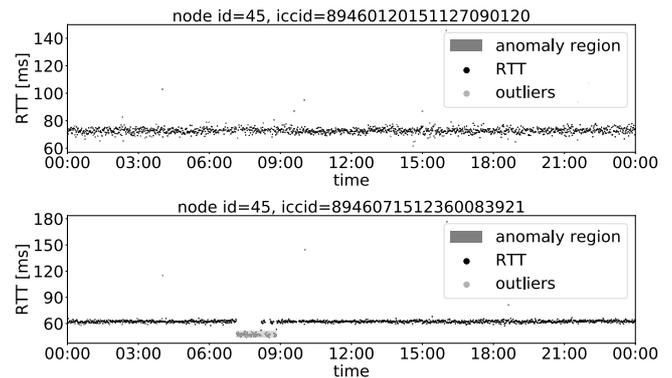

Fig. 13. Anomaly occurs only at one interface (bottom image) of the same node.

*Effects of mobile broadband measurement system on the results.* Uncovering the role of seemingly unrelated system design decisions on KPI values is one of the key observations we arrived to, as we tested RICERCANDO on MONROE data. For instance, after significant amounts of meta-data started arriving from MONROE nodes we discovered that RTT exhibits occasional spikes (going above 5X the usual value) in-terspersed with lost ping packets. Further analysis with Rapid Exploration tools uncovered correlation between the observed anomaly and the node resource utilisation spikes, indicating potential executions of CPU-heavy experiments. Consequently, our suggestion to include experiment execution information in the metadata was implemented by the MONROE team, which later allowed us to pinpoint a particular experiment that resulted in the observed RTT behaviour. This is just one example where the measurement system, in this case through heavy resource usage by an experiment, resulted in anomalous measurements. The impact of the background traffic on RTT measurements via DRX mode toggling is another example of the coupling of the measurement methodology and the recorded result, and is explained in Section V.

All interfaces sent the ping probe to the same destination host IP of a server at Karlstad University, Sweden. We noticed that the nodes located in Norway and Sweden often had mean RTT of the ping probe from 40 to 60 ms, while it was not uncommon for the nodes in countries far from destination host server to encounter mean RTT close to 100 ms. This observation reveals that the anomalies emerge because of relative changes in feature values at one node rather than by comparing the absolute difference of feature values among distant nodes.

*MBB measurement data analysis requires multidisciplinary expertise.* While we were already aware of the need for interdisciplinary expertise at the time we laid out plans for RICERCANDO, this need became even more evident as we progressed with development. First, MBB data is often analysed by computer networking domain experts. The need for expertise in data mining, in particular in data representation, statistical analysis, and geographical data analysis proved crucial and the data mining part of our team got several enquiries to help with other projects' data analysis issues. The two fields, data mining and computer networking, are seldom directly collaborating, and it is our hope that RICERCANDO results might facilitate this collaboration. Second, even when the general knowledge of networking is present, MBB measurement data mining requires in-depth knowledge of latest practices in broadband networks' implementation. Such knowledge is often available only with a close collaboration with relevant industrial players. Specifically, our identification of the DRX-related anomaly would not be possible without close collaboration with an industry professional experienced with LTE networks.

## VII. CONCLUSIONS

In this paper we presented RICERCANDO – an MBB measurement data mining toolkit developed in close collaboration of networking and machine learning experts. RICERCANDO goes beyond the existing tools by allowing rapid iterative visual analysis and rigorous advanced data mining of MBB data. In this paper we present a few use cases demonstrating the usability of the framework for anomaly detection and root cause explanation. Although the framework was designed primarily for the analysis of data collected in MONROE testbed, its usability is by no means restricted to a particular dataset. We have already harnessed RICERCANDO for mining MBB measurement data gathered by the Slovenian Agency for Telecommunications (AKOS) with the goal of inferring Internet neutrality violations in Slovenia. RICERCANDO toolbox has a great potential to assist commercial telcos and government regulators with monitoring and understanding MBB traffic, and we invite interested parties to download RICERCANDO[9], adapt it to their needs, enrich it with additional functionalities, and further contribute towards improved MBB measurement data analysis and understanding.

## ACKNOWLEDGEMENTS

The authors would like to thank Prof Fabio Ricciato for his guidance during the planing, execution, and writing about the work presented in this paper; the data mining team from the University of Ljubljana, including Jernej Kernc, Vesna Tanko, and Anže Starič for their contributions to RICERCANDO; to Janez Sterle for his help with the explanation of the observed network anomalies; and to David Modic for his feedback on an earlier draft of this paper.

---

[9] http://github.com/ivek1312/ricercando/

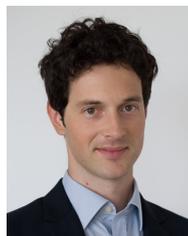

**Veljko Pejovic** received his PhD in computer science from the University of California, Santa Barbara, USA in 2012 on the topic of resource-efficient wireless communication for rural areas. From 2012 to 2014 Dr Pejovic worked as a research fellow at the University of Birmingham, UK in the area of mobile computing and sensing. His work on modelling users' movement and communication behaviour from mobile call records has won the 2013 Orange Data for Development Challenge, while his work on on developing machine learning models of interruptibility based on sensor data resulted in the best paper nomination at the 2014 ACM UbiComp conference. Currently, he is an assistant professor at the Faculty of Computer and Information Science, University of Ljubljana, Slovenia, where he works on mobile sensing and resource-efficient mobile computing.

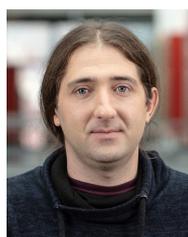

**Ivan Majhen** received the Diploma degree in computer and information science from the University of Ljubljana, Slovenia. He has been working on WiFi-Direct implementation for Arduino platform at Jozef Stefan Institute, Slovenia in 2014. In 2018, he joined Computer Communications Laboratory at Faculty of computer and information science, University of Ljubljana. He is currently focusing on wireless sensors for human vital signs monitoring.

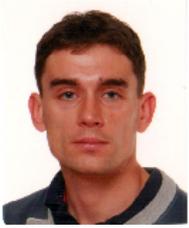

**Miha Janež** received the PhD in computer science from the University of Ljubljana, Slovenia in 2012 on the topic of circuit layout design. He is an assistant at the Faculty of Computer and Information Science, University of Ljubljana, Slovenia. His research interests include the design of wireless sensor networks using accessible hardware components and the analysis of the data collected in large networks.

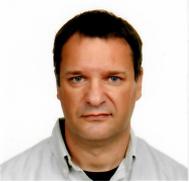

**Blaž Zupan** heads the bioinformatics lab at University of Ljubljana and is an Associate Professor at the Baylor College of Medicine in Houston. His research has focused on constructive induction, machine learning and epistasis approaches to reconstruction of gene networks, large-scale data fusion, and data visualizations. He believes that crafting simple tools that anybody can use to understand data is essential to advancements of humanity and democracy. His lab is developing Orange, a fully open-source, ever evolving data mining suite with a visual programming environment. He also enjoys writing scripts for YouTube videos to explain data science, and preparing courses that introduce data science.